\documentclass[aps,pre,showpacs,floatfix]{revtex4}%
\usepackage{amsmath}
\usepackage{graphicx}
\usepackage{color}
\usepackage{bm}
\usepackage{verbatim}
\usepackage{amsfonts}
\usepackage{amssymb}
\usepackage[tight,sf]{subfigure}
\usepackage{psfrag}

\newcommand{\romd}{{\text{d}}}

\newcommand{\VECn}{{\boldsymbol{n}}}
\newcommand{\VECX}{{\boldsymbol{X}}}

\newcommand{\TENG}{{\text{\textsf{G}}}}
\newcommand{\TENS}{{\text{\textsf{S}}}}
\newcommand{\TENX}{{\text{\textsf{X}}}}

\begin{document}
\title{Interface-Mediated Interactions: Entropic Forces of Curved Membranes}
\author{Pierre Gosselin$^{1}$, Herv\'{e} Mohrbach$^{2}$, and Martin Michael
M\"{u}ller$^{2}$}
\affiliation{$^{1}$Institut Fourier, UMR 5582 CNRS-UJF, Universit\'{e} Grenoble I, BP74,
38402 St Martin d'H\`{e}res, France }
\affiliation{$^{2}$Equipe BioPhysStat, ICPMB-FR CNRS 2843, Universit\'{e} Paul
Verlaine-Metz; 1, boulevard Arago, 57070 Metz, France}
\date{\today}

\begin{abstract}
Particles embedded in a fluctuating interface experience forces and torques
mediated by the deformations and by the thermal fluctuations of the medium.
Considering a system of two cylinders bound to a fluid membrane we show that
the entropic contribution enhances the curvature-mediated repulsion between
the two cylinders.
This is contrary to the usual attractive Casimir force in
the absence of curvature-mediated interactions. For a large distance between
the cylinders, we retrieve the renormalization of the surface tension of a
flat membrane due to thermal fluctuations.

\end{abstract}


\maketitle


\section{Introduction}

Particles bound to an interface may interact via forces and torques of two
distinct physical origins. One contribution to these so-called
interface-mediated interactions is purely geometric and results from the
deformations caused by the particles. Since the interface is also a thermally
fluctuating medium, embedded particles may also interact through a
fluctuation-induced interaction. The associated entropic force is an example
of the more general phenomenon of Casimir forces between objects placed in a
fluctuating medium. In its original formulation, two uncharged conducting
plates were predicted to attract each other due to the quantum electromagnetic
fluctuations of the vacuum \cite{CASIMIR}. In a soft matter context,
fluctuation-mediated forces were, for instance, studied for objects immersed
in a fluid near its critical point \cite{FisherdeGennes,Hertleinetal,Kresch}
or attached to a fluid interface \cite{LiKardar,Dietrich,NoruzifarOettel}.

Interface-mediated forces have also received intense attention recently due to
their possible relevance in biological processes: membrane-mediated
interactions could aid cooperation of proteins in the biological membrane and
complement the effects of direct van der Waals' or electrostatic forces
\cite{MARTINNATURE}. Theoretical studies of this problem have typically
considered particles on quasi-planar fluid membranes
\cite{GOULIAN,GOLESTANIAN,PARKLUBENSKY,KIMNEUOSTER,FOURNIER,WEIKL,
MarkusCasimir,RoyaCasimir} neglecting the intrinsic nonlinearity of the
underlying (ground state) shape equation. Some systems, especially those with a 
symmetry, have been studied on a nonlinear level without taking into account any 
fluctuations \cite{BisBis}. 

In this paper, we investigate interface-mediated interactions in their full
generality on a \emph{curved} geometry \emph{including entropic contributions}
for the specific problem of two parallel cylinders bound to the same side of a
membrane. The ground state of this problem and thus the forces at zero
temperature induced by the membrane were studied in Refs.~\cite{MARTIN2005,MARTIN2007} via stress and torque tensors and in Refs.~\cite{MARTIN2007,MkrtchyanIngChen} via energy minimization. 
The method employed here to include thermal fluctuations is based on the
calculation of the free energy of the system in a semi-classical
approximation, where Gaussian fluctuations around the curved ground state are
computed. To this end we introduce a new parametrization for the fluctuation
variables which is possible due to the translational symmetry of the membrane.
The force can then be obtained by deriving the free energy with respect to the
distance between the cylinders.


\section{The model}

\begin{figure}[ptb]
\begin{center}
\includegraphics*[width=0.47\textwidth, bb=51 380 557 620]{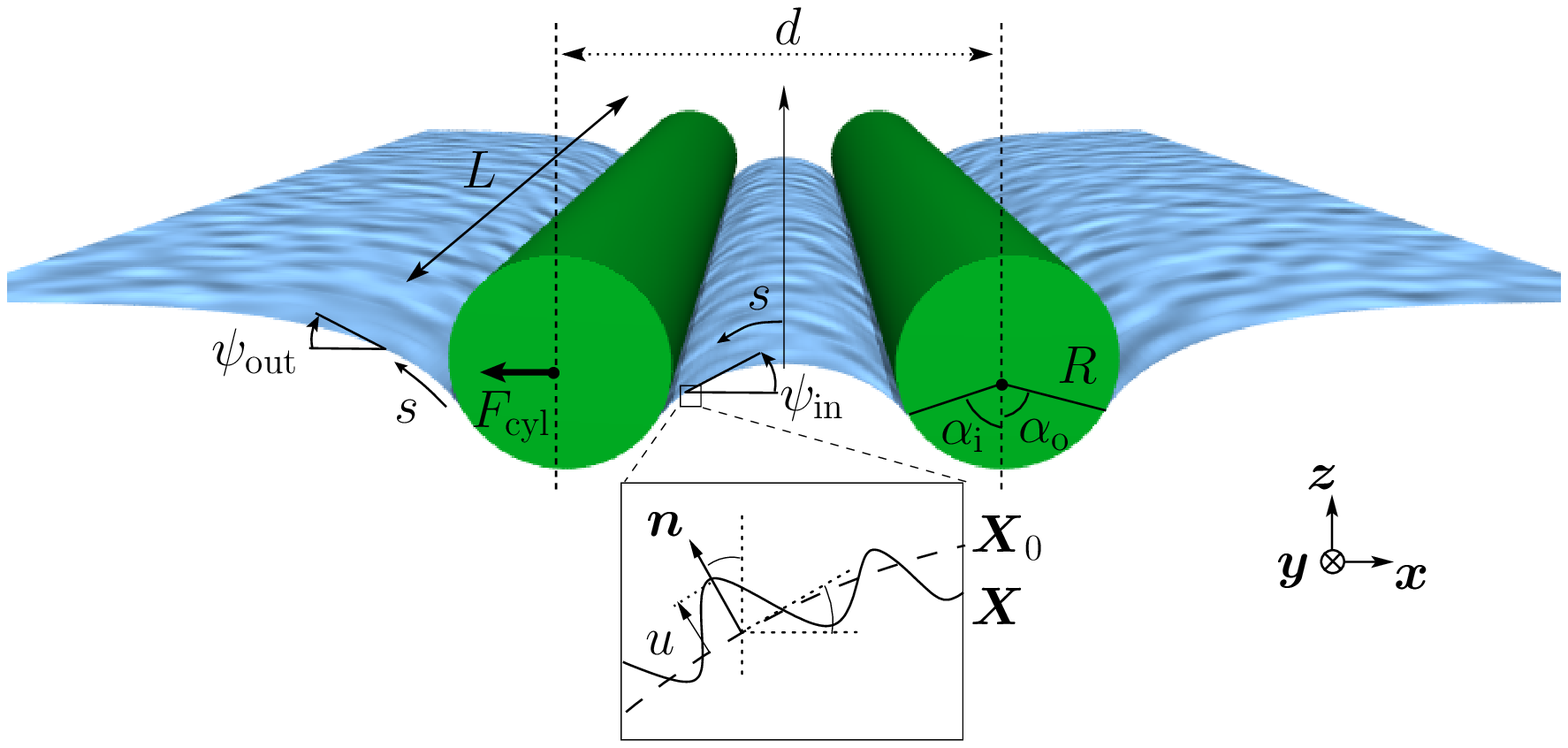}
\end{center}
\caption{(Color online) Two parallel cylinders on the same side of a fluid membrane with
fixed wrapping angle $\alpha_{\text{c}}=\alpha_{\text{i}}+\alpha_{\text{o}%
}=120^{\circ}$. The membrane fluctuates around its ground state profile
${\boldsymbol{X}}_{0}$. The corresponding fluctuation variable $u$ gives the
distance between ${\boldsymbol{X}}_{0}$ and the actual profile
${\boldsymbol{X}}$ measured along the normal ${\boldsymbol{n}}$ of the ground
state membrane (see inset).}%
\label{fig:system}%
\end{figure}

We first start by exposing the problem and shortly retrieve the ground state
configuration which will be the starting point for the computation of the
thermal fluctuations. Consider two identical cylinders of length $L$ and
radius $R$ bound to one side of the membrane, parallel to the $y$ axis and
separated by a distance $d$ (see Fig.~\ref{fig:system}).
In the limit of large $L/R$, boundary effects at the ends of the cylinders can
be neglected and the profile can be decomposed into the following parts: an
inner section between the cylinders, two outer sections that become flat for
$x\rightarrow\pm\infty$, and two bound sections in which the cylinder and the
membrane are in contact with each other. The contact area is given by
$\alpha_{\text{c}}R L$ where $\alpha_{\text{c}}$ is the wrapping angle (see
Fig.~\ref{fig:system}). The value of $\alpha_{\text{c}}$ depends on the
physical situation considered:
$i)$ the cylinder either has a finite adhesion energy $w$ per area so that
$\alpha_{\text{c}}$ is determined via an adhesion balance at the contact lines
or $ii)$ only a given part of the cylinder surface is adhering strongly to the
membrane so that $\alpha_{\text{c}}$ is fixed. The ground state of case $i)$,
which was studied numerically in detail in Ref.~\cite{MkrtchyanIngChen},
displays a phase diagram which is more complicated than in case $ii)$. In
order to avoid complications already at the ground state level we will thus
focus on case $ii)$ in the following by setting $\alpha_{\text{c}}%
=\alpha_{\text{o}}+\alpha_{\text{i}}=\text{const}$, where $\alpha
_{\text{o}/\text{i}}$ is the contact angle between the cylinder and the
outer/inner membrane. The shape of the bound parts is prescribed by the
geometry of the attached cylinder, whereas the profiles of the free membrane
sections are determined by solving the nonlinear shape equation which results
from the minimization of the Helfrich Hamiltonian \cite{Canham,Helfrich}
\begin{equation}
H=\int_{\Sigma}\text{d}A\,\left(  \sigma+\frac{\kappa}{2}K^{2}\right)  \;,
\label{H}%
\end{equation}
where $\Sigma$ is the surface of the free membrane and $\text{d}A$ is the
infinitesimal area element. In this functional, $\sigma$ denotes the surface
tension, $\kappa$ the bending rigidity, and $K$ the local curvature of the
membrane. At zero temperature the profile obeys translational symmetry along
the $y$ axis. It is thus convenient to introduce the angle-arc length
parametrization $\psi(s)$ where $s$ is the arc length and $\psi$ the angle
between the $x$ axis and the tangent to the profile. In this parametrization
the curvature is given by $K=\pm\text{d}\psi/\text{d}s$ [$+$ for the inner
region and $-$ for the outer ones (see again Fig.~\ref{fig:system})]. The shape
equation of the surface can be written as $\lambda^{2}\text{d}^{2}%
\psi/\text{d}s^{2}-\eta\sin\psi=0$ with $\lambda=\sqrt{\kappa/\sigma}$ the
reference length scale. The dimensionless quantity $\eta$ is defined as
$\eta=f_{x}/\sigma$, where $f_{x}$ is the force per length $L$ of the cylinder
at every point of the membrane (which is constant and horizontal on each
membrane section). Using the stress tensor \cite{MARTIN2007}, the zero
temperature force on the left cylinder is given by the simple expression
$F_{\text{cyl}}^{(0)}=\sigma(\eta-1)L$. The outer section exercises a pulling
force $-\sigma L$ on the left cylinder, the force of the inner section is
$\sigma\eta L$. Later on, we will see how the values of the forces will be
renormalized by thermal fluctuations. For the total force without fluctuations
one only has to determine the value of $\eta$ which is an implicit function of
$d.$ To do so, one first solves the shape equation which admits different
solutions $\psi_{\text{in}}(s)$ and $\psi_{\text{out}}(s)$ (expressed in terms
of elliptic Jacobi functions) corresponding to the inner and outer sections
and depending on the boundary conditions at each cylinder. The value of $\eta$
in the inner section for any given $\alpha_{\text{i}}$ is determined
implicitly by the requirement $\psi_{0}\equiv\psi_{\text{in}}(s_{0}%
)=\alpha_{\text{i}}$ where $s_{0}$ is the arc length between the mid-line and
the contact point on the cylinder of the inner membrane. Then $s_{0}$ is also
implicitly determined by the relation
\begin{equation}
\frac{d}{2} -R\sin\alpha_{\text{i}}=\int_{0}^{s_{0}}\text{d}s\cos\psi\; .
\label{eq:arclength}%
\end{equation}
The torque balance equation at equilibrium,
\begin{equation}
K_{\text{i}}-K_{\text{o}} - \frac{R}{\lambda^{2}} (\eta\cos{\alpha_{\text{i}}%
}-\cos{\alpha_{\text{o}}})=0 \; , \label{eq:torquebalance}%
\end{equation}
where $K_{\text{o}/\text{i}}$ is the contact curvature in the outer/inner
region, fixes the individual values $\alpha_{\text{o}}$ and $\alpha_{\text{i}%
}$. Solving the torque equation, values of $\eta$, $\alpha_{\text{o}}$,
$\alpha_{\text{i}}$ for a given $d$ can now be determined numerically. For the
case under consideration $\eta<1$ is an increasing function with the distance
$d$ so that the cylinders always repel each other \cite{MARTIN2007}. This
justifies that we have restricted our discussion to parallel cylinders.
Indeed, every deviation from parallelism would directly be compensated by a
counteracting torque.


\section{Thermal fluctuations}


\subsection{The fluctuation operator $H^{(2)}$\label{sec:fluctuationoperator}}

With the knowledge of the zero temperature profile, it is now possible to
compute the entropic force. We first focus on the \emph{inner} section and set
$\psi (s) \equiv\psi_{\text{in}} (s)$. The position vector including fluctuations can
then be written as ${\boldsymbol{X}}(s,y) = {\boldsymbol{X}}_{0}(s) + (-u
\sin{\psi} , 0 , u \cos{\psi})$ where $u(s,y)$ is the membrane fluctuation in
the normal direction and ${\boldsymbol{X}}_{0}(s)$ the position vector of the
zero temperature profile (see inset of Fig.~\ref{fig:system}).

For small temperature the Helfrich Hamiltonian~(\ref{H}) can be expanded
$H=H_{0}+H^{(2)}$ to second order in $u$, where $H_{0}$ is the ground state
energy. The contribution $V(d)$ of the fluctuations to the free energy is then
given by:
\begin{equation}
V(d)=\beta^{-1}\ln\int\text{D}u\,e^{-\beta H^{(2)}}\;,
\label{FL}%
\end{equation}
where $\beta^{-1}=k_{B}T$. The fluctuation operator $H^{(2)}$ can be
determined by expressing the Helfrich Hamiltonian~(\ref{H}) with the help of
the parametrization ${\boldsymbol{X}}(s,y)$. 
%
One obtains (see appendix~\ref{app:fluctuationoperator})
\begin{eqnarray}
H^{(2)}  &  = & \int \left[\frac{\dot{\psi}^{4}u^{2}}{2}+\dot{\psi}\ddot{\psi}%
u_{s}u+\Big( \frac{1}{\lambda^{2}}+\frac{3\dot{\psi}^{2}}{2} \Big)\frac{u_{s}^{2}}
{2}+\dot{\psi}^{2}u_{ss}u \right.
\nonumber\\
&&  \left. +\frac{u_{ss}^{2}+u_{yy}^{2}}{2}+u_{ss}u_{yy} 
+ \Big(\frac{1}{\lambda^{2}} -\frac{\dot{\psi}^{2}}{2}\Big)\frac{u_{y}^{2}}{2}\right]\text{d}s\text{d}y
\;, \label{H2}
\label{eq:fluctuationoperator}
\end{eqnarray}
where $u$ was assumed to satisfy periodic boundary conditions. 
The domain of integration is $-L/2<y<L/2$ and $-s_{0}<s<s_{0}.$ The thermal
contribution to the force of the inner section is in principal given by
$F_{\text{in}}^{\text{fl}}=\partial V(d)/\partial d$. However, thermal
fluctuations also induce a rotation of the cylinders to maintain the torque
balance. For small membrane curvatures the actual values of $\eta$ as well as
$s_{0}$ and $\psi_{0}$ differ only slightly from their zero temperature
values. Solving the arc length and torque equations (\ref{eq:arclength}) and
(\ref{eq:torquebalance}) for small deviations $\delta\eta$, $\delta s_{0}$,
$\delta\psi_{0}$, one can see that the inner thermal force must be corrected
by a prefactor, $i.e.$, $Z(\psi_{0},\eta)F_{\text{in}}^{\text{fl}}$. It turns
out that $Z(\psi_{0},\eta)$ does not vary much from unity and is thus
disregarded here. For a large curvature of the inner membrane this
approximation would in principle break down even though a general change of
the behavior is not expected.

The computation of Eq.~(\ref{FL}) for every value of separation $d$ is very
difficult as $H^{(2)}$ has no known eigenvalues and eigenfunctions. To
circumvent this problem we focus on the two limiting cases, the
\emph{quasi-flat} and the \emph{highly curved regime} and propose an
interpolating formula for intermediate separations.


\subsection{The quasi-flat regime\label{sec:quasiflatregime}}

Let us first consider the \emph{quasi-flat regime}, \emph{i.e.}, the regime of
very large $d/\lambda$ for which the membrane can be considered as flat except
at the cylinders. In this case $\eta\approx1$ and $\partial\psi_{0}/\partial
d\approx0$. In Eq.~(\ref{FL}) $u$ can be expanded in Fourier modes
$u(s,y)=\sum_{q,n}u_{n,q}\exp(i\pi ns/s_{0})\exp(i\pi qy/L)$ with $n$ and $q$
two integers. An implicit cut-off $\Lambda$ of the order of the inverse of the
membrane thickness ($a\sim5\text{nm}$) is assumed. The number of modes along
the $y$ -/$s$-direction are given respectively by $N=2\pi\Lambda L$ and
$M=4\pi\Lambda s_{0}$ ($2s_{0}\approx d$ being the arc length of the inner
part). Since the field $u(s,y)$ is dimensional, the measure
of the partition function is $\text{D}u\equiv\prod_{y,s}\mu^{-1}%
\text{d}u(s,y)$ with $\mu$ an arbitrary length scale which
disappears from the expression of the force. 

To compute the energy contribution~(\ref{FL}) in the quasi-flat regime, we decompose 
Eq.~(\ref{H2}) in two parts:
\begin{eqnarray}
H^{(2)} & = &\int\left[\frac{u_{ss}^{2}+u_{yy}^{2}}{2}+u_{ss}u_{yy}+\frac{1}{\lambda
^{2}}\frac{u_{y}^{2}}{2}\right]\text{d}s\text{d}y 
\nonumber \\
&& \, + \, \int \left[\frac{\dot{\psi}^{4}u^{2}}%
{2}+\dot{\psi}\ddot{\psi}u_{s}u+\Big(\frac{1}{\lambda^{2}}+\frac{3\dot{\psi}^{2}%
}{2}\Big)\frac{u_{s}^{2}}{2}+\dot{\psi}^{2}u_{ss}u-\frac{\dot{\psi}^{2}}{2}%
\frac{u_{y}^{2}}{2}\right]\text{d}s\text{d}y\;
\label{eq:H2decomposition}
\end{eqnarray}
and rewrite $H^{(2)}$ in Fourier components according to this
decomposition:
\begin{equation}
H^{(2)}=4dL\beta\mu^{2}\sum_{m,n}u\left(  -m\right)  \left[
\text{\textsf{G$^{-1}$}}(-m,n,q)+{\text{\textsf{X}}}(-m,n,q)\right]  u\left(
n\right)  \; ,
\end{equation}
where the $u (n) $ are the Fourier components of $u$, with 
$u (-m)  =\bar{u} (m)$.
The terms \textsf{G$^{-1}$} and $\TENX$ correspond to the Fourier
transform of the two terms arising in the decomposition~(\ref{eq:H2decomposition}). Namely, one has:%
\begin{equation}
\text{\textsf{G$^{-1}$}}(-m,n,q)=\delta_{mn}\left\{\left[\Big(\frac{\pi n}{d}\Big)^{2} 
+ \Big(\frac{\pi q}{L}\Big)^{2}\right]^{2}+\frac{1}{\lambda^{2}}\left[\Big(\frac{\pi n}{d}\Big)^{2} 
+ \Big(\frac{\pi q}{L}\Big)^{2}\right]\right\}
\; ,
\end{equation}
which is the diagonal propagator of the flat case and
\begin{align}
{\text{\textsf{X}}}(-m,n,q)  &  =\frac{1}{d}\int_{-\frac{d}{2}}^{\frac{d}{2}%
}{\text{d}}s\;\Bigg[\frac{1}{2}\dot{\psi}^{4}+\frac{3}{2}\left(  \ddot{\psi
}^{2}+\frac{1}{\lambda^{2}}\dot{\psi}^{2}\right) \nonumber\\
&  -\!\!\frac{5\pi^{2}}{4}\left(  \frac{mn}{d^{2}}-\frac{q^{2}}{L^{2}}\right)
\dot{\psi}^{2}\Bigg]\exp{\Big[\frac{i\pi(n-m)s}{d}}\Big]
\end{align}
the non-diagonal matrix due to curvature corrections. 
These results allow to compute the path integral~(\ref{FL}) with the help of the usual formula 
for the integral of a quadratic weight:
\begin{align}
\beta V  &  =\ln\int\text{D}u\,e^{-\beta H^{(2)}}\;=\ln\sqrt{\det\left[
4dL\beta\mu^{2}\left(  \text{\textsf{G$^{-1}$}}+\text{\textsf{X}}\right)  \right]  }\nonumber \\
&  =\frac{1}{2}\operatorname{Tr}\ln[4dL\beta\mu^{2}(\text{\textsf{G$^{-1}$}%
}+\text{\textsf{X}})]\nonumber \\
&  =\frac{1}{2}\operatorname{Tr}\ln{(4dL\beta\mu^{2}\text{\textsf{G$^{-1}$}}%
)}+\sum_{k\geq1}\frac{(-1)^{k-1}}{2k}\operatorname{Tr}({\TENG%
}{\text{\textsf{X}}})^{k}\;.\;\;\;\; \label{eq:perturbativeexp}%
\end{align}
The first term of expression~(\ref{eq:perturbativeexp}) is just the free
energy of the flat case:
\begin{align}
\frac{1}{2}\sum_{q,n}\ln\Bigg(4dL\beta\mu^{2}   &  \bigg\{ \bigg[\frac
{1}{2}\bigg(\frac{2\pi n}{d}\bigg)^{2}+\bigg(\frac{2\pi q}{L}\bigg)^{2}%
\bigg]^{2}\nonumber\\
&  \!\!\!\!+\,\frac{1}{2\lambda^{2}}\bigg[\bigg(\frac{2\pi n}{d}%
\bigg)^{2}+\bigg(\frac{2\pi q}{L}\bigg)^{2}\bigg]\bigg\}\Bigg)\;.\;\;\;\;
\end{align}
The second term is the perturbation correction (with $\tilde{\text{\textsf{G}%
}}(n,q):=\TENG(-n,n,q)$): 
\begin{eqnarray}
\sum_{k=1}^{\infty }\frac{(-1)^{k-1}}{2k}\operatorname{Tr}\left( \TENG\text{\textsf{X}}\right) ^{k}
=\sum_{k=1}^{\infty }\frac{\left( -1\right) ^{k-1}}{2k}\sum_{q}
\sum_{n_{i},\atop \sum n_{i}=0} && \!\text{\textsf{X}} ( -n_{1},n_{2},q)
\tilde{\TENG}(n_{2},q) \text{\textsf{X}}( -n_{2},n_{3},q) \tilde{\TENG}(n_{3},q )
\nonumber \\
&& \ldots \,\text{\textsf{X}}( -n_{k-1},n_{k},q) \tilde{\TENG}(n_{k},q)
\; .
\label{eq:perturbation2ndterm}
\end{eqnarray}
A careful inspection shows that to the leading order $1/d$
the sum in Eq.~(\ref{eq:perturbation2ndterm}) is dominated by terms where the
propagator $\TENG$ is singular, that is at $\mathbf{(}n,q)\sim0$. As the
term $\TENX(-m,n,q)$ is regular at the origin, we can approximate the series
Eq.~(\ref{eq:perturbation2ndterm}) by keeping only the contributions of the form 
$\TENX(0,0,0)\tilde{\TENG}(n,q)$. This dominant contribution can be 
resummed as:
\begin{equation}
\sum_{k=1}^{\infty}\frac{(-1)^{k-1}}{2k}\operatorname{Tr}\left(
\TENG\text{\textsf{X}}\right)  ^{k}\simeq\frac{1}{2}\sum_{q}%
\ln\big[1+\sum_{n}\text{\textsf{X}}(0,0,0)\tilde{\TENG}(n,q)\big]
\end{equation}
yielding a correction to the force:
\begin{equation}
 \beta\delta F=\frac{1}{2}\sum_{q}\frac{\frac{\partial}{\partial d}\sum
_{n}\text{\textsf{X}}(0,0,0)\tilde{\TENG}(n,q)}{1+\sum
_{n}\text{\textsf{X}}(0,0,0)\tilde{\TENG}(n,q)}
\; .
\label{eq:deltaF}
\end{equation}%
\begin{figure}[ptb]
\psfrag{Ftil}{$-\tilde{F}^{\text{fl}}/L$}
\par
\begin{center}
  \includegraphics*[width=0.45\textwidth]{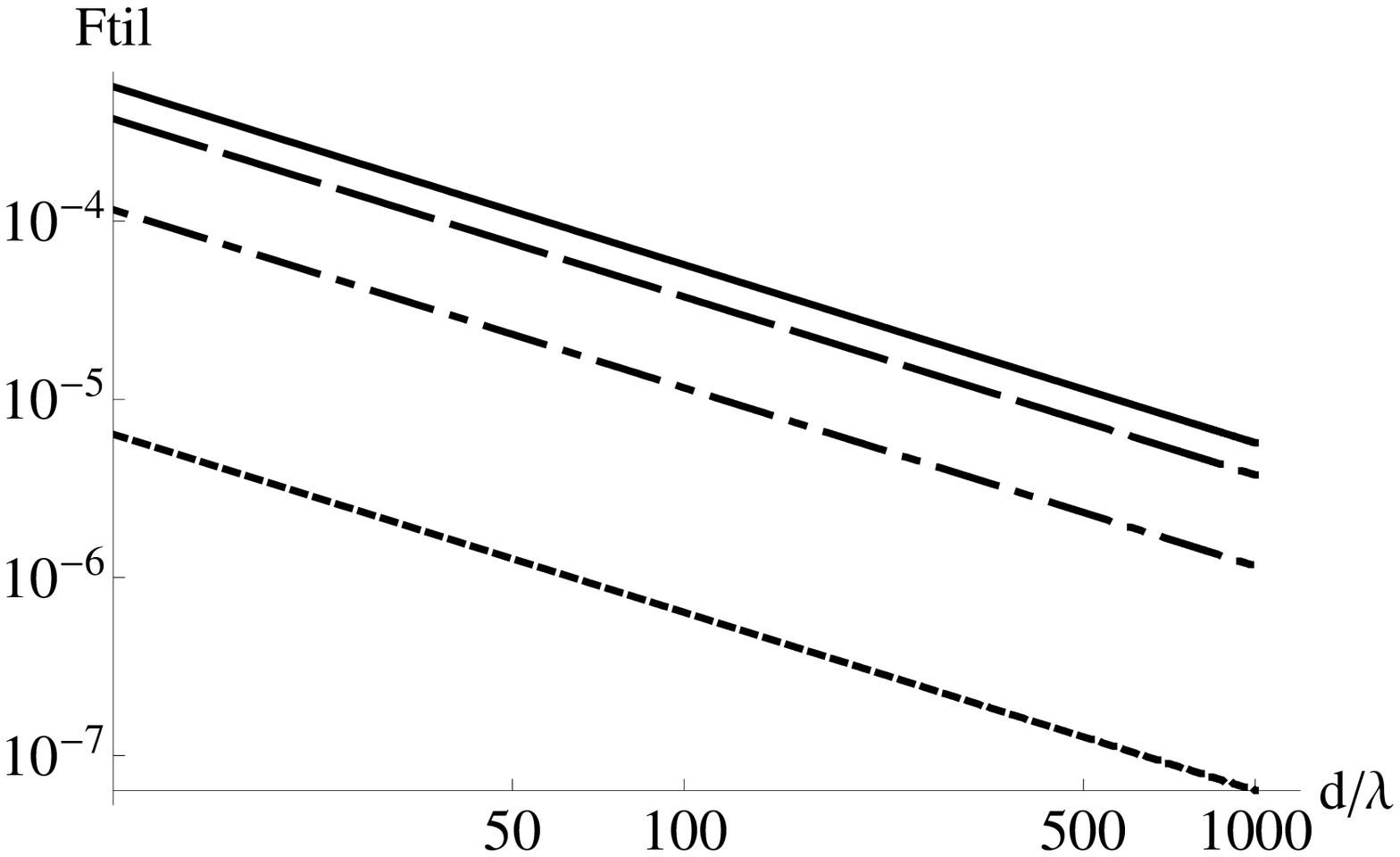}
\end{center}
\par
\caption{Scaled total thermal force $-\tilde{F}^{\text{fl}}/L=-2 \pi \beta F^{\text{fl}} / (L\Lambda^{2})=2\pi\mathcal{A}\lambda/d$ on the left cylinder in the 
quasi-flat regime as a function of separation $d/\lambda$ for $\alpha_{\text{c}}=10^{\circ}$ (short
dashes), $\alpha_{\text{c}}=45^{\circ}$ (dashed-dotted line), $\alpha
_{\text{c}}=90^{\circ}$ (long dashes), and $120^{\circ}$ (solid line).}%
\label{fig:largedlimit}%
\end{figure}%
In the limit of large $d$ and $L$ the sums can be approximated by integrals. A
lengthy but straightforward calculation, which is sketched in appendix~\ref{app:forcelargedlimit}, leads to the following expression of
the thermal force of the inner section
\begin{equation}
\frac{\beta F_{\text{in}}^{\text{fl}}}{L\Lambda^{2}}\approx-\frac{1}{2\pi^{2}%
}-\mathcal{A}\,\frac{\lambda}{d}+\text{o}\left(  \frac{\lambda}{d}\right)
\;\text{\ for }d/\lambda>>1\label{Fquasiflat}
\end{equation}
with $\mathcal{A}=\frac{\mathcal{B}}{2\pi^{2}\left(  \Lambda\lambda\right)
^{2}}\int_{0}^{1}\frac{1}{1+x^{2}}\frac{dx}{1+\frac{\mathcal{B}}{\pi
\Lambda\lambda}\frac{1}{x}\arctan(\frac{1}{x})}>0$, where $\mathcal{B}%
=\frac{16\mathcal{C}\left(  4\mathcal{C}^{2}+3\mathcal{C}+9\right)  }{3\left(
\mathcal{C}+1\right)  ^{3}}$ and $\mathcal{C}=\tan^{2}\left(  \psi
_{0}/4\right)  $. Since $F_{\text{in}}^{\text{fl}}$ is negative, it
contributes to the repulsion between the cylinders. For $d\rightarrow\infty$,
$F_{\text{in}}^{\text{fl}}/L\approx-\Lambda^{2}/(2\pi^{2}\beta)$ corresponds
to the entropic part of the intrinsic tension as found in Ref.~\cite{FOURNIER2} and
denoted by $\tau\equiv\tau_{\text{in}}=\sigma-\Lambda^{2}/(2\pi^{2}\beta)$.
Obviously, the outer section pulls with an intrinsic tension of opposite sign
$\tau_{\text{out}}=-\tau_{\text{in}}$ on the left cylinder. The total thermal
force $F^{\text{fl}}=-L\Lambda^{2}\mathcal{A}\lambda/(\beta d)$ in the
quasi-flat regime is thus always repulsive just as the zero temperature force
$F_{\text{cyl}}^{(0)}$ (see Fig.~\ref{fig:largedlimit}).
It should be noted that even for $\eta$ close to one the membrane is not completely flat due to the nonvanishing contact angle $\psi_0$.
Only for $\psi_0 = 0$ is the membrane completely flat and $F^{\text{fl}}$ equals zero.
To recover the usual attractive Casimir behavior one has to go beyond the $1/d$ expansion.%


\subsection{The highly curved regime}

At small separations the membrane is highly curved if the scaled ground state
force $\eta$ is close to zero \cite{MARTIN2007}. This \emph{highly curved
regime} is accessible only for large wrapping angles $\alpha_{\text{c}}$: for
example, choosing the radius of the cylinders such that $R/\lambda=1$, then
$d/\lambda\geq2$. If the two cylinders are in contact, \emph{i.e.},
$d/\lambda=2$, one obtains $\eta=0.93$ for $\alpha_{\text{c}}=45^{\circ}$
which implies that the membrane is still rather flat. For $\alpha_{\text{c}%
}=120^{\circ}$ one is already close to the highly curved regime since
$\eta=0.11$.


\subsubsection{Change of variable}

To calculate the thermal force in this regime, consider a---zero energy
cost--- rigid translation of amplitude unity in both $x$ and $z$ directions.
This translation of the membrane as a whole can be decomposed into normal and
tangential components which are combinations of $\cos\psi$ and $\sin\psi$.
Now, owing to the property that tangential fluctuations leave $H$ invariant
\cite{GUVEN}, it is clear that individually $\cos\psi$ and $\sin\psi$ are zero
modes of $H^{(2)}$. As they are also zero modes of the operator $\text{d}%
^{2}/\text{d}\psi^{2}+1$, we can write $H^{(2)}$ in the form $H^{(2)}=\frac
{1}{2}\int[\dot{\psi}^{2}(\text{d}^{2}/\text{d}\psi^{2}+1)u]^{2}+u\tilde
{H}(y)u\text{d}y$ where $\tilde{H}(y)$ is a differential operator acting on
the variable $y$ only. This form of $H^{(2)}$ and the fact that from the
ground state we have $\text{d}s=\text{d}\psi/K(\psi)$ with $K(\psi
)=\sqrt{2[1-\eta\cos(\psi)]}/\lambda$, suggests a change of variables between
$s\rightarrow\psi=\psi(s)$. In this way, the $s$ dependence is eliminated in
the benefit of the new variable of integration, $\psi$, and%
\begin{align}
H^{(2)}  &  =\frac{1}{2}\int \Big\{ K(\psi)^{3}[(\frac{\text{d}^{2}}{\text{d}\psi
^{2}}+1)u]^{2}-2K(\psi)\frac{\partial u_{yy}}{\partial\psi}\frac{\partial
u}{\partial\psi} \nonumber\\
& +[K(\psi)^{-1}/\lambda^{2}-K(\psi)/2]u_{y}^{2}+u_{yy}^{2}\Big\}
\romd\psi \romd y\;, \label{H2P}%
\end{align}
where $u(s,y)$ has been replaced by $u(\psi,y)$ \cite{FootnoteG}. The domain
of integration is now $-L/2<y<L/2$ and $-\psi_{0}<\psi<\psi_{0}$. This new
formulation is the clue to compute the Gaussian functional integral~(\ref{FL})
since it relies only on the angle variable $\psi$. Using the Fourier transform
$u(\psi,y)=\sum_{q,n}\tilde{u}_{n,q}\exp(i\pi n\psi/\psi_{0})\exp(i\pi qy/L)$
with $n$ and $q$ two integers, we can write $H^{(2)}=\sum_{m,n,q}\tilde
{u}_{m,q}{\text{\textsf{S}}}(m,n,q)\tilde{u}_{n,q}$ where
\begin{align}
{\text{\textsf{S}}}(m,n,q)  &  =2\psi_{0}\Bigg\{ \frac{\big[ ( \frac{\pi
m}{\psi_{0}})^{2}-1\big] \big[ ( \frac{\pi n}{\psi_{0}})^{2}-1\big] }%
{2}a_{n+m}\nonumber\\
&  - \, \left(  \frac{\pi q}{L}\right)  ^{2}\left(  \frac{\pi^{2} n m}%
{\psi_{0}^{2}}\right)  b_{n+m} + \frac{1}{2} \left(  \frac{\pi q}{L}\right)
^{4}c_{n+m}\nonumber\\
&  + \,\frac{1}{2} \left(  \frac{\pi q}{L}\right)  ^{2}\left(  \frac{c_{n+m}%
}{\lambda^{2}}-\frac{b_{n+m}}{2}\right)  \Bigg\}
\end{align}
with
\begin{subequations}
\begin{align}
a_{k}  &  = \frac{1}{2\psi_{0}}\int_{-\psi_{0}}^{\psi_{0}} K(\psi
)^{3}\mathcal{W}_{k}(\psi) {\text{d}}\psi\; ,\\
b_{k}  &  = \frac{1}{2\psi_{0}}\int_{-\psi_{0}}^{\psi_{0}} K(\psi
)\mathcal{W}_{k}(\psi) {\text{d}}\psi\; , \quad\text{and}\\
c_{k}  &  = \frac{1}{2\psi_{0}}\int_{-\psi_{0}}^{\psi_{0}} K(\psi
)^{-1}\mathcal{W}_{k}(\psi) {\text{d}}\psi\; ,
\end{align}
\label{eq:akbkck}
\end{subequations}
where $\mathcal{W}_{k}(\psi):=\exp( -i\pi k\psi/\psi_{0})$.


\subsubsection{The cut-off problem}

In the usual Fourier decomposition $u(s,y)=\sum_{q,n}u_{n,q}\exp(i\pi
ns/s_{0})\exp(i\pi qy/L)$ an implicit cut-off $\Lambda$ is assumed with
$N=2\pi\Lambda L$ and $M=4\pi\Lambda s_{0}$ the number of modes along the
$y$-/$s$-direction, respectively (see Sec.~\ref{sec:quasiflatregime}). A short
inspection shows that for a finite $M$ in the $s$ space there is an infinite
number of modes in the $\psi$ space. Actually, an expansion in the $\psi$
variable of a function $F(\psi)$ has to be equal to its expansion in $s$
space:
\begin{equation}
F\left(  \psi\left(  s\right)  \right)  =\sum_{n}\tilde{F}_{n}\exp\left(
\frac{i\pi n\psi\left(  s\right)  }{\psi_{0}}\right)  =\sum_{n}F_{n}%
\exp\left(  \frac{i\pi ns}{s_{0}}\right)
\; .
\end{equation}
The relation $F_{n}=\sum_{p} A_{np}\tilde{F}_{p}$ connects both kind of Fourier coefficients. 
It can be found by expanding $\psi(s)$ in its own Fourier components 
$\psi_{k}$. Actually, the expansion of the exponential of a sum of
exponentials is given by the Jacobi Anger formula which leads to
\begin{widetext}
\begin{equation}
A_{np} = 
\left[ \prod_{k=1}^{\infty }J_{0}\left( \tilde{\psi}_{k}^{\left( p\right) }\right) \right] 
\left( \delta_{np}+\sum_{l=1}^{\infty } \sum_{k_{i}>0,\atop i=1...l}
\sum_{m_{k_{i}}\neq 0,\atop \sum_{i=1}^{l}k_{i}m_{k_{i}}=n-p}
\prod_{i=1}^{l}\frac{J_{m_{k_{i}}}\left( \tilde{\psi}_{k_{i}}^{\left( p\right) }\right) }%
{J_{0}\left( \tilde{\psi}_{k_{i}}^{\left( p\right) }\right) }\right)  
\; ,
\label{eq:Anp}
\end{equation}
\end{widetext} 
with $\tilde{\psi}_{k}^{\left(  p\right)  }=\frac{2\pi
ip\psi_{k}}{\psi_{0}}$ where $J_{k}$ are Bessel functions of the first kind \cite{Abramowitz}.
Consequently, the path integral of $H^{(2)}$ over $N$ modes $u_{n,q}$ in the
$s$ space should involve an infinite number of modes $\tilde{u}_{n,q}$ in the
$\psi$ space. To clarify this point, consider the Gaussian weight in the $s$ space \footnote{For 
the sake of simplicity, we suppress the variable $q$ which does not play any role in the following 
argument and write $S_{nm}$ for $\TENS(n,m,q)$.}, $\exp(\sum_{-N<n,m<N}%
u_{-m}S_{nm}u_{n})$. We will see in the next section that it can be approximated by a diagonal 
quadratic form in the $\psi$ space,  $\exp(\sum_{p}\tilde{u}_{-p}\tilde{S}_{pp}\tilde{u}_{p})$. 
The corresponding coefficients $\tilde{S}_{pp}$ are obviously given by the change of 
variables~(\ref{eq:Anp})
\begin{equation}
\tilde{S}_{pp}=\left(  \sum_{-N<n,m<N}S_{nm}A_{mp}A_{np}\right)  \;.
\end{equation}
However, a careful analysis shows an exponential decrease of the coefficients
$\tilde{S}_{pp}$ for $p>M$ which is faster the closer $\eta$ to $0$, but
relatively slow for $\eta$\ close to $1$. This relies on the fact that for
small $\eta$ the coefficient $A_{np}$ can be shown to be equal to
$\delta_{np}+\eta pC(-1)^{n-p}/(n-p)^{3}\left(  1-\delta_{np}\right)  $ with
$C=\frac{4}{\pi^{2}}\frac{s_{0}^{2}}{\lambda^{2}}\frac{\sin\psi_{0}}{\psi_{0}%
}$. For $\eta=0$, $A_{np}=\delta_{np}$ and all $\tilde{S}_{pp}=0$ for $p>M$;
the same cut-off $\Lambda$ can thus be implemented in $s$ and $\psi$ space.
Therefore, as long as we stay close to $\eta=0$, we consider a constant cut-off
$\Lambda$. This condition has to be relaxed when $\eta$ goes to one. In
Sec~\ref{sec:interpolation} we will propose an interpolating formula between
the two regimes $\eta\approx0$ and $\eta\approx1$.


\subsubsection{Interaction energy for the highly curved regime}

In principle we could do a perturbative expansion of the same kind as in the
quasi-flat case. A careful inspection shows that the small expansion parameter
is $\eta$\ so that the perturbative contributions due to the off-diagonal
matrix elements ${\text{\textsf{S}}}(m,n,q)$ are negligible for $\eta\approx0$. 
Therefore, $\beta V\approx
2^{-1}\operatorname{Tr}\ln{\text{\textsf{D}}}$ with ${\text{\textsf{D}}}%
\equiv{\text{\textsf{S}}}(-\underline{n},\underline{n},\underline{q})$ given
explicitly by ${\text{\textsf{S}}}=2\psi_{0}L[(\underline{n}^{2}-1)^{2}%
a_{0}+(2b_{0}\underline{n}^{2}+d_{0})\underline{q}^{2}+c_{0}\underline{q}%
^{4}]$ with the notations $\underline{n}=\pi n/\psi_{0}$ and $\underline
{q}=\pi q/L.$ The various coefficients are functions of $d$. They are given by
Eqs.~(\ref{eq:akbkck}) with $k=0$ and $d_{0}=c_{0}/\lambda^{2}-b_{0}/2$
and can be explicitly evaluated in terms of Jacobi elliptic functions.
Interestingly, for $\eta=0$, we have $a_{0}=2b_{0}/\lambda^{2}=4c_{0}%
/\lambda^{4}=4/(\sqrt{2}\lambda^{4})$ and ${\text{\textsf{S}}}(\underline
{m},\underline{n},\underline{q})=\delta_{\underline{m},-\underline{n}}
8\sqrt{2}\psi_{0}L\lambda^{-4}\{(\underline{n}^{2}-1)^{2}+(\underline
{q}\lambda)^{2}[\underline{n}^{2}+(\underline{q}\lambda)^{2}/4]\}$ which, for
$\psi_{0}=\pi$, corresponds to the propagator obtained in
\cite{Fourniertubule} for membrane tubules. In the limit of large $L$, the sum
over $q$ can be replaced by an integration $\sum\nolimits_{q}\rightarrow
\int\nolimits_{-\Lambda}^{\Lambda}$ and we obtain with $n^{\prime}%
=\sqrt{\underline{n}^{2} - 1}$
\begin{align}
\frac{\beta V(d)}{L\Lambda}  &  =\frac{1}{\pi}\sum_{n=1}^{M}\ln[2\psi
_{0}L\beta\mu^{2}(n^{\prime4}a_{0}+2\Lambda^{2} n^{\prime2} b_{0}
+c_{0}\Lambda^{4})]\nonumber\\
&  -\frac{2n^{\prime}c_{+}}{b}[\arctan(\frac{b\Lambda+n^{\prime}c_{-}%
}{n^{\prime}c_{+}}) +\arctan(\frac{b\Lambda-n^{\prime}c_{-}}{n^{\prime}c_{+}%
})]\nonumber\\
&  -[4\Lambda-\frac{n^{\prime}c_{-}}{b}\ln(\frac{an^{\prime2}+b\Lambda
^{2}+2\Lambda n^{\prime}c_{-}}{an^{\prime2}+b\Lambda^{2}-2\Lambda n^{\prime
}c_{-}})] \;. \label{VD}%
\end{align}
The coefficients $a,b,c_{\pm}$ are given by $a=\sqrt{a_{0}},$ $b=\sqrt{c_{0}}$
and $c_{\pm}=[(\sqrt{a_{0}c_{0}}\pm b_{0})/2]^{1/2}$. The calculation of the
entropic force $F_{\text{in}}^{\text{fl}}=\partial V(d)/\partial d$ is now
straightforward but must be done with caution. First, all coefficients $a,b,c$
as well as $\psi_{0}$ are implicit functions of $d.$ Second, the
differentiation with respect to $d$ which is a continuous variable must be
done at constant $\Lambda$ even though, at first sight, the number of modes
$M=2\pi\Lambda d$ is proportional to $d$. But $M$ is actually the integer part
of $2\pi\Lambda d$ and is thus insensitive to an infinitesimal change of $d.$
Moreover, Eq.~$\left(  \ref{VD}\right)  $ is strictly speaking only valid in
the highly curved regime where $\eta\approx0.$ To go to larger distances one
has to take into account the variation of the number of modes (due to our
change of variable $s\rightarrow\psi)$ as well as the off-diagonal elements of
${\text{\textsf{S}}}(m,n,q)$ which could be computed perturbatively.


\subsection{Interpolation formula for all separations\label{sec:interpolation}%
}

Instead of adjusting exactly the discrete number of modes with $d$ (which is
in fact impossible)$,$ we choose to introduce a two parameter function
$g(p,\alpha)=(p\,s_{0}/\lambda)^{\eta/\alpha}$, such that the Fourier modes
$n$ in Eq.~$\left(  \ref{VD}\right)  $ must be replaced by $\widetilde
{n}=n/g(p,\alpha)$ everywhere. This ansatz takes into account the growing
number of modes with $d$ in an approximate but controlled manner. The
parameters $p$ and $\alpha$ have to be chosen such that in the highly curved
regime $g(p,\alpha)\sim1$ whereas in approaching $\eta\approx1,$ the force
should correspond to Eq. (\ref{Fquasiflat}). This ansatz has thus a second
advantage: like a variational procedure would also do, it allows to
approximate the perturbative contributions which are very hard to compute.
Therefore, the introduction of $\widetilde{n}$ is a way to interpolate between
the two regimes $\eta\approx0$ and $\eta\approx1$. Taking all this into
account and replacing the $\sum\nolimits_{n}$ by an integral, the thermal
force $F_{\text{in}}^{\text{fl}}$ on the left cylinder reads
\begin{align}
\frac{2\pi\beta F_{\text{in}}^{\text{fl}}}{L\Lambda^{2}}=  &  \frac{2}{\pi
}(\frac{\partial s_{0}}{\partial d}-\frac{s_{0}\partial g}{g\partial d}%
)+\frac{\sqrt{2}}{x}(\sqrt{x-y}U+\sqrt{x+y}V)\nonumber\\
&  +\frac{1}{2gx^{2}}[(x^{2}+g^{2}y)U-g^{2}\sqrt{x^{2}-y^{2}}V]\ln
A\nonumber\\
&  -\frac{g}{x^{2}}(\sqrt{x^{2}-y^{2}}U+yV)(\arctan B_{+}+\arctan
B_{-})\nonumber\\
&  +\frac{V}{g}(\arctan D_{+}+\arctan D_{-}) \label{FC}%
\end{align}
where $x=(a_{0}c_{0}^{3})^{1/2}$ and $y=b_{0}c_{0}.$ We also introduced the
notations $A=\frac{x+g^{2}+g\sqrt{2\left(  x-y\right)  }}{x+g^{2}%
-g\sqrt{2\left(  x-y\right)  }}$, $B_{\pm}=\frac{\sqrt{2x}\pm g\sqrt{x-y}%
}{g\sqrt{x+y}}$ and $D_{\pm}=\frac{g\sqrt{2}\pm\sqrt{x-y}}{\sqrt{x+y}}$ as
well as $U=\frac{2^{1/2}}{\pi}\{s_{0}\frac{\partial}{\partial d}%
(x-y)^{1/2}-(x-y)^{1/2}\frac{\partial}{g\partial d}(s_{0}g)\}$ and
$V=\frac{2^{1/2}}{\pi}\{s_{0}\frac{\partial}{\partial d}(x+y)^{1/2}%
-(x+y)^{1/2}\frac{\partial}{g\partial d}(s_{0}g)\}$. \ Asking that the large
$d/\lambda$ limit of Eq.~(\ref{FC}) is given by Eq.~(\ref{Fquasiflat}) we
obtain $\alpha=5$ and $p=\delta\exp\left(  -2\pi^{2}\mathcal{A}\frac{\lambda
}{d}\right)  $ where the constant of integration $\delta$ is the only free
parameter determined below.

\begin{figure}[ptb]
\psfrag{Ftil}{$-\tilde{F}^{\text{fl}}/L$}
\par
\begin{center}
  \includegraphics*[width=0.45\textwidth]{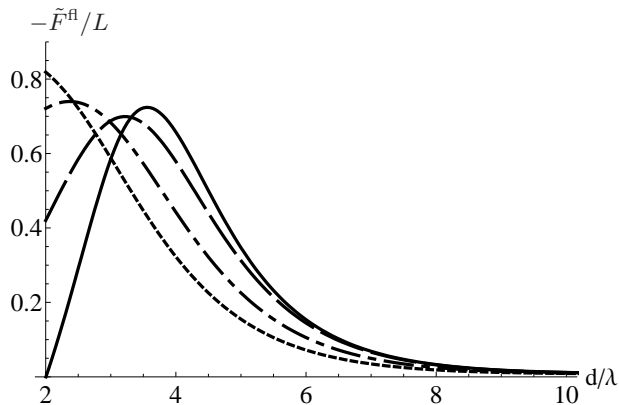}
\end{center}
\caption{Scaled total thermal force $-\tilde{F}^{\text{fl}}/L:=-2\pi\beta
F^{\text{fl}}/(L\Lambda^{2})=-[2\pi\beta F_{\text{in}}^{\text{fl}}%
/(L\Lambda^{2})+1/\pi]$ on the left cylinder as a function of separation
$d/\lambda\geq2$ for $R/\lambda=1$ and $\alpha_{\text{c}}=10^{\circ}$ (short
dashes), $\alpha_{\text{c}}=45^{\circ}$ (dashed-dotted line), $\alpha
_{\text{c}}=90^{\circ}$ (long dashes), and $120^{\circ}$ (solid line).}%
\label{fig:thermalforce}%
\end{figure}

\begin{figure}[ptb]
\psfrag{Ftin}{$-\tilde{F}_{\text{in}}^{\text{fl}}/L$}
\psfrag{A}{$\delta=10^{-1}$} \psfrag{B}{$\delta=10^{-2}$}
\psfrag{C}{$\delta=10^{-3}$} \psfrag{D}{$\delta\le 10^{-6}$}
\par
\begin{center}
  \includegraphics*[width=0.45\textwidth]{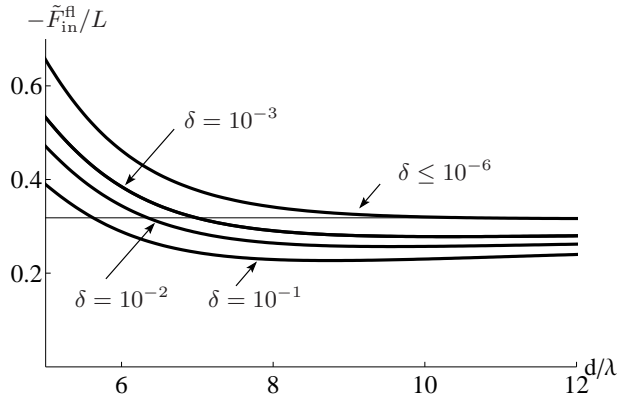}
\end{center}
\caption{Scaled thermal force $-\tilde{F}_{\text{in}}^{\text{fl}}%
/L:=-2\pi\beta F_{\text{in}}^{\text{fl}}/(L\Lambda^{2})$ on the left cylinder
as a function of separation $d/\lambda\geq2$ for $R/\lambda=1$, $\alpha
_{\text{c}}=120^{\circ}$ and fitting parameters $\delta=10^{-1}, 10^{-2}, 10^{-3}$, and
$\le 10^{-6}$. The first three curves reach the limit $1/\pi$ 
at $d/\lambda\rightarrow\infty$ from below. All curves with $\delta \le 10^{-6}$ are 
identical for the given resolution of the figure and give a monotous unique force.}%
\label{fig:fixingdelta}%
\end{figure}

Inserting numerical values in Eq.~(\ref{FC}) one finds that the thermal force
$F_{\text{in}}^{\text{fl}}$ exerted on the left cylinder by the inner part of
the membrane is negative. It thus enhances the curvature-mediated repulsion
between the cylinders. As the outer freely fluctuating membrane exerts a
constant force $F_{\text{out}}^{\text{fl}}/L=\Lambda^{2}/(2\pi^{2}\beta)$
which does not compensate $F_{\text{in}}^{\text{fl}}$ completely, the total
thermal force $F^{\text{fl}}$ is repulsive for all separations (see
Fig.~\ref{fig:thermalforce}).
\begin{figure}[ptb]
\begin{center}
\subfigure[]{\includegraphics*[width=0.45\textwidth]{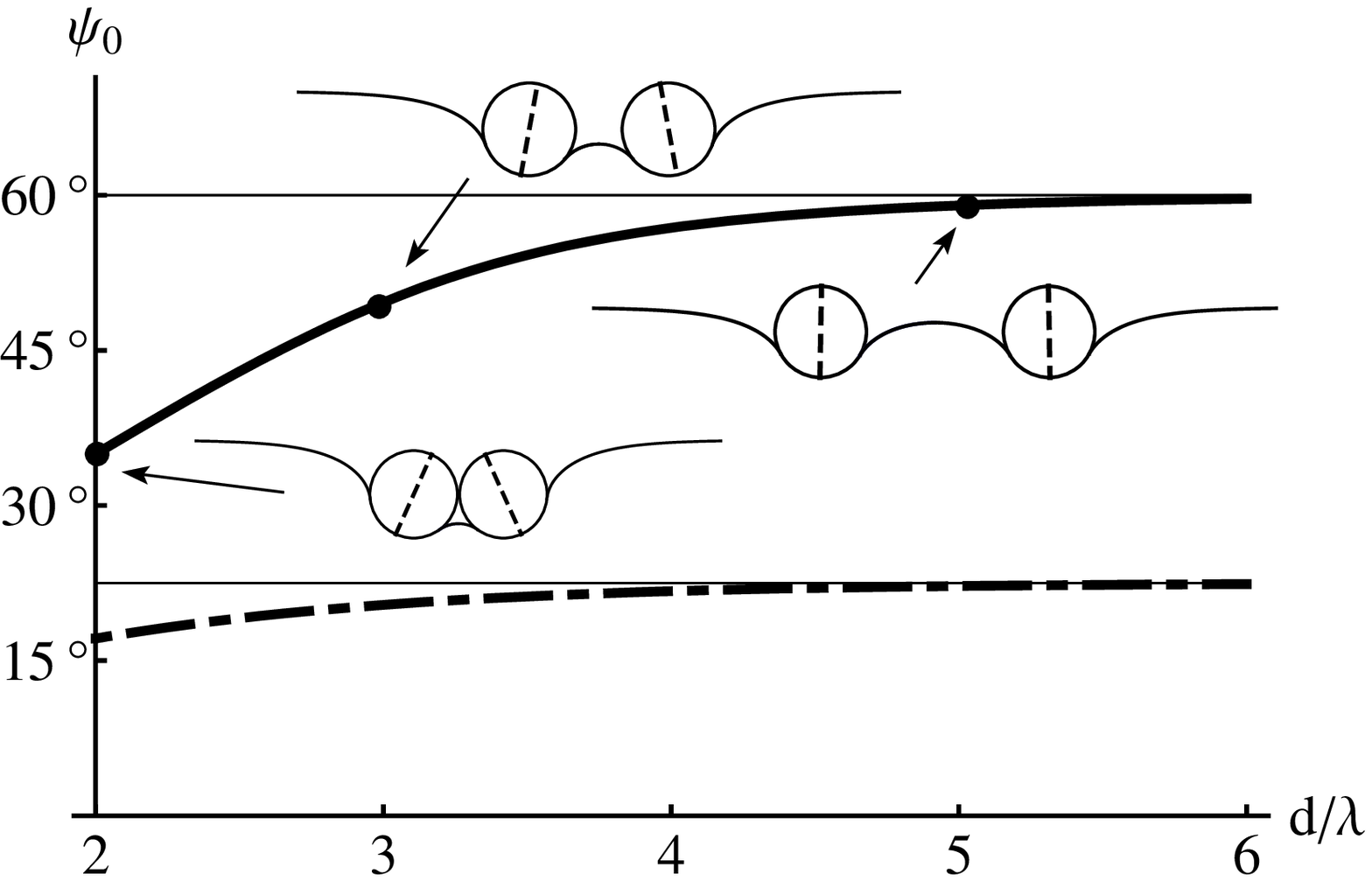}}
\hfill
\subfigure[]{\includegraphics*[width=0.45\textwidth]{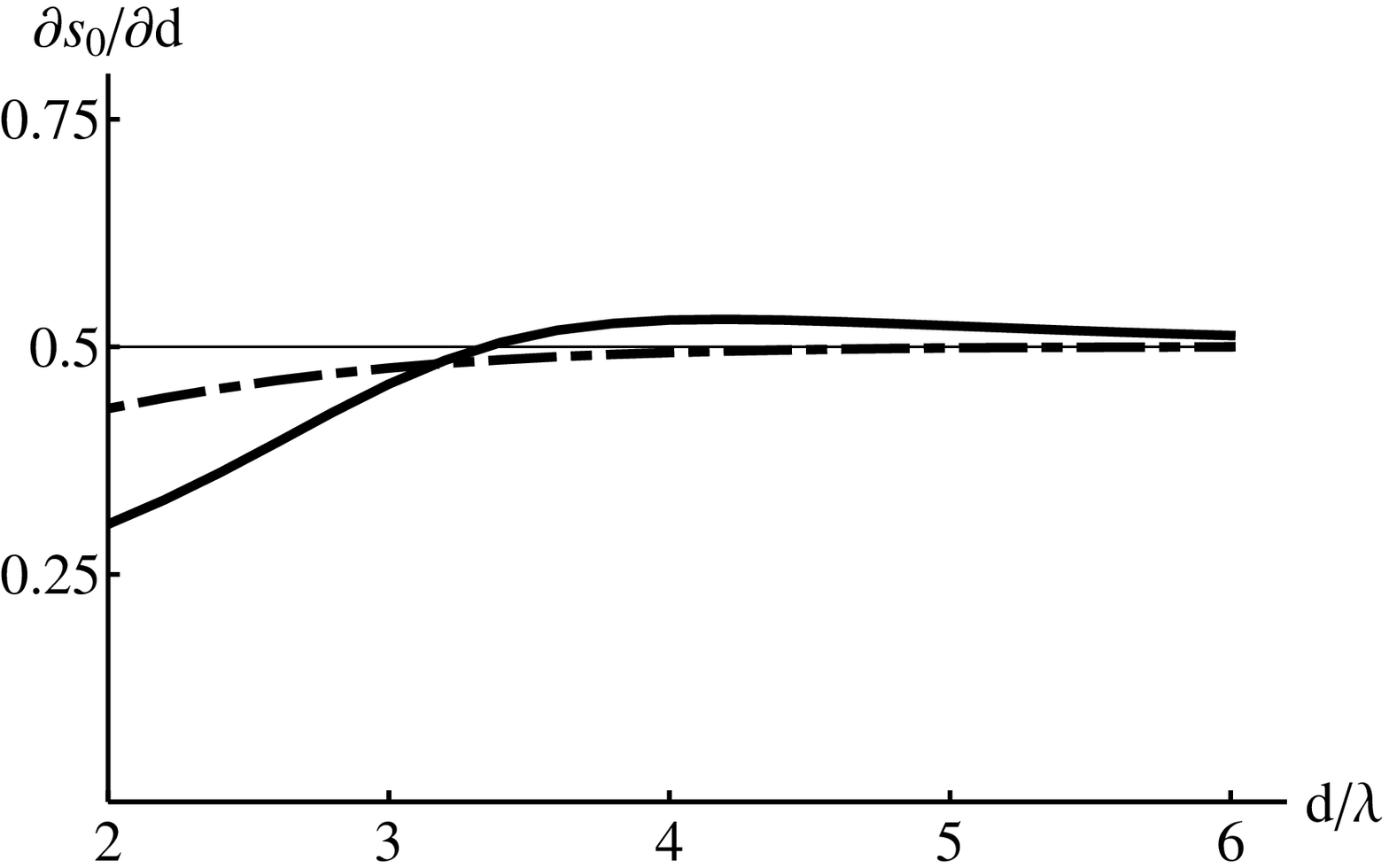}}
\end{center}
\caption{(a) Inner contact angle $\psi_{0}$ and (b) derivative of the arc
length $\partial s_{0}/\partial d$ as a function of $d/\lambda$ for
$R/\lambda=1$ and $\alpha_{\text{c}}=45^{\circ}$ (dashed-dotted line) and
$120^{\circ}$ (solid line). The thin solid lines correspond to the respective
large $d$ limits.}%
\label{fig:psi0s0}%
\end{figure}
The curve of $-F^{\text{fl}}$ shows two characteristic trends: at short
separations $d/\lambda\approx2,$ the force increases with $d$. This is due to
a fast rotation of the cylinders for an infinitesimal change in $d$ (see
Fig.~\ref{fig:psi0s0}) which implies that the length $2s_{0}$ of the inner
membrane stays almost unchanged, \emph{i.\ e.}, $\partial s_{0}/\partial d$ is
small. The membrane is thus more under tension and thermal fluctuations are
strongly reduced.
Since $\partial s_{0}/\partial d$ increases with $d$, the force grows until it
reaches a maximum value at $\partial s_{0}/\partial d\approx1/2$. For larger
separations, $\psi_{0}$ and $\partial s_{0}/\partial d$ stay constant and  
the force $-F_{\text{in}}^{\text{fl}}$ should decrease in a monotonous manner until it
tends to the constant value $L\Lambda^{2}/(2\pi^{2}\beta)$ in the limit
$d/\lambda\rightarrow\infty$. Actually, this monotonous decrease of the
force---expected on physical grounds---can be exploited to fix the fitting 
parameter $\delta$: as shown in Fig.~\ref{fig:fixingdelta}, $\delta$ has to be 
set to a value smaller than $10^{-6}$. 

To summarize, the total force per length $F_{\text{cyl}}/L=(F_{\text{cyl}%
}^{(0)}+F^{\text{fl}})/L$ on the left cylinder is
\begin{equation}
F_{\text{cyl}}/L=-\sigma(1-\eta)+F_{\text{in}}^{\text{fl}}/L+\Lambda^{2}%
/(2\pi^{2}\beta) \; .
\end{equation}
Since this force is always negative, there is no equilibrium position beside
the limit $d/\lambda\rightarrow\infty$, where $F_{\text{cyl}}\rightarrow0$.


\section{Conclusion} 
By developing a new approach for the computation of the
free energy for a system of two cylinders bound on the same side of a
membrane, we could evaluate the corrections caused by the thermal fluctuations
to the repelling zero temperature force. It was found that this contribution
in the section between the cylinders strongly depends on the membrane
curvature. The calculated thermal force is always repulsive. This effect differs 
from the attractive Casimir force which arises from the reduction of the number 
of internal modes with respect to the outer ones where the ground state is
identical everywhere. This is obviously not the case here since the zero
temperature shapes of the inner and outer sections are different: even though
the number of modes in the inner section is smaller than in the outer ones,
their fluctuations are strongly enhanced on a curved background and thus
always dominate the fluctuations of the outer less curved region. Non-trivial
membrane geometries as the one presented here are in fact ubiquitous in
nature. The approach of this paper is sufficiently general to calculate the
physical properties of other systems with highly curved ground states.

This research was supported in part by the National Science Foundation under
Grant No.\ NSF PHY05-51164.


\begin{appendix}

\section{Derivation of the fluctuation operator $H^{(2)}$ (see 
Eq.~(\ref{eq:fluctuationoperator})) \label{app:fluctuationoperator}}

To rewrite the Helfrich Hamiltonian~(1) in terms of the parametrization $\VECX (s,y)$, one 
has to replace the area element $\romd A$ and the curvature $K$ in (1) using the functions 
$u(s,y)$, $\psi (s)$, and their derivatives \cite{DifferentialGeometry}. First, one needs to 
determine the components of the (symmetric) metric tensor $g_{ab}$ ($a, b\in\{s,y\}$) 
\footnote{Note that $\partial\VECX_0/\partial s = (\cos{\psi},0,\sin{\psi})$.}: 
\begin{eqnarray}
  g_{ss} & = & \left(\frac{\partial \VECX}{\partial s}\right)^2  
    = (u \dot{\psi} - 1)^{2} + u_{s}^{2} \; ,
  \nonumber \\
  g_{sy}´& = & \frac{\partial \VECX}{\partial y}\cdot \frac{\partial \VECX}{\partial s} 
    = u_{y} u_{s} \; , \quad \text{and}
  \nonumber \\
  g_{yy} & = & \left(\frac{\partial \VECX}{\partial y}\right)^2  = 1+u_{y}^{2}
  \; ,
\end{eqnarray}
where we have introduced  the notations $\dot{\psi
}\equiv\partial\psi/\partial s$, $u_{s} \equiv\partial u/\partial s$ and
$u_{y}\equiv\partial u/\partial y$. This allows to calculate the area element
${\text{d}} A = \sqrt{g} \,{\text{d}} s {\text{d}} y$ with $\sqrt{g}%
=\sqrt{\det{(g_{ab})}}=\sqrt{(u \dot{\psi} - 1)^{2} (1+u_{y}^{2}) + u_{s}^{2}%
}$. The components of the extrinsic curvature tensor $K_{ab}$ are:
\begin{eqnarray}
  K_{ss} & = &   \frac{\partial \VECX}{\partial s}\cdot \frac{\partial \VECn}{\partial s}    
  = \frac{1}{\sqrt{g}} 
    \left[ -\dot{\psi} + 2 u \dot{\psi}^{2} - u_{ss} - u^{2} \dot{\psi}^{3} - u
u_{s} \ddot{\psi} + (u u_{ss} - 2 u_{s}^{2}) \,\dot{\psi} \right]
  \nonumber \\
  K_{sy} & = & \frac{\partial \VECX}{\partial y}\cdot \frac{\partial \VECn}{\partial s}    
  = \frac{1}{\sqrt{g}} 
    \left[ (-1 + u \dot{\psi}) \, u_{sy} - \dot{\psi} u_{y} u_{s} \right] \; , \quad \text{and}
  \nonumber \\
  K_{yy} & = & \frac{\partial \VECX}{\partial y}\cdot \frac{\partial \VECn}{\partial y}    
  = \frac{1}{\sqrt{g}} (-1 + u \dot{\psi}) \, u_{ss}
  \; .
\end{eqnarray}
The contraction of $K_{ab}$ with the 
metric yields the curvature $K = \sum K_{ab}g^{ab}$ where $g^{ab}$ is the inverse of 
the metric, $i.e.$, $g^{ss}=g_{yy}/g$, $g^{sy}=-g_{sy}/g$, and $g^{yy}=g_{ss}/g$. 
Inserting the expressions for $\romd A$ and $K$ into $H$ one identifies
\begin{eqnarray}
H^{(2)}  &  =\int[\frac{\dot{\psi}^{4}u^{2}}{2}+\dot{\psi}\ddot{\psi}%
u_{s}u+(\frac{1}{\lambda^{2}}+\frac{3\dot{\psi}^{2}}{2})\frac{u_{s}^{2}}%
{2}+\dot{\psi}^{2}u_{ss}u\nonumber\\
&  +\frac{u_{ss}^{2}+u_{yy}^{2}}{2}+u_{ss}u_{yy}+(\frac{1}{\lambda^{2}}%
-\frac{\dot{\psi}^{2}}{2})\frac{u_{y}^{2}}{2}]\text{d}s\text{d}y\; .
\end{eqnarray}


\section{Derivation of Eq.~(\ref{Fquasiflat}) \label{app:forcelargedlimit}}

Let $X (0)  = \TENX (0,0,0)$ and start with expression (\ref{eq:deltaF})
\begin{equation}
\beta \delta F=\frac{1}{2}\sum_{q}\frac{\frac{\partial}{\partial d}\sum
_{n} X (0) \tilde{\TENG} ( n,q ) }{1+\sum_{n} X (0)
\tilde{\TENG} ( n,q )}
\; .
\label{deltaf}
\end{equation}
Note that, according to their definition, 
$\frac{\partial}{\partial d} X (0)  =-\frac{1}{d} X (0)$ and $\frac{\partial}{\partial
d}\sum_{n}\tilde{\TENG}( n,q )  =\frac{2}{d}\sum_{n}\frac{\lambda^{2}\left(
\frac{2\pi n}{d}\right)  ^{2}}{\left[  \left(  \frac{2\pi n}{d}\right)
^{2}+\left(  \frac{2\pi q}{L}\right)  ^{2}\right] ^{2}}$. Thus, a continuous
approximation for the sum (which is valid for large $d$, given the high number
of modes) yields for large separations:
\begin{align*}
\frac{\partial}{\partial d}\sum_{n} X (0)  \tilde{\TENG} ( n,q )
&  =-\frac{ X ( 0 )  }{\pi}\lambda^{2}\int_{\frac{2\pi}{d}}^{\Lambda
}\left[  \frac{1}{\left(  n^{2}+q^{2}\right)  }-\frac{2n^{2}}{\left(
n^{2}+q^{2}\right)  ^{2}}\right]  \romd n\\
&  \simeq-\frac{X ( 0 )  }{\pi}\lambda^{2}\Lambda\frac{q^{2}%
-\frac{2\pi}{d}\Lambda}{\left(  \Lambda^{2}+q^{2}\right)  \left[  \left(
\frac{2\pi}{d}\right)  ^{2}+q^{2}\right]  }%
\end{align*}
and by the same token:%
\[
1+\sum_{n}X (0)  \tilde{\TENG} ( n,q )  = 1+\frac{\lambda^{2}}{\pi} 
X (0) d \int_{\frac{2\pi}{d}}^{\Lambda} \frac{1}{n^{2}+q^{2}} \,\romd n
\simeq 1+\frac{\lambda^{2}}{\pi}\frac{X (0)  d}{q}\arctan\left(
\frac{q}{\frac{2\pi}{d}+\frac{q^{2}}{\Lambda}}\right)\,.
\]
As a consequence Eq.~(\ref{deltaf}) becomes in the continuum approximation:%
\[
\beta\delta F = -\frac{1}{d}\frac{L}{2\pi}\int_{\frac{2\pi}{d}}^{\Lambda}%
\frac{\frac{X\left(  0\right)  d}{\pi}\lambda^{2}\Lambda\frac{q^{2}-\frac
{2\pi}{d}\Lambda}{\left(  \Lambda^{2}+q^{2}\right)  \left[  \left(  \frac
{2\pi}{d}\right)  ^{2}+q^{2}\right]  }}{1+\frac{\lambda^{2}}{\pi} \frac{X (0)  d}{q}%
\arctan\left(  \frac{q}{\frac{2\pi}{d}+\frac{q^{2}%
}{\Lambda}}\right)  } \romd q \; .
\]
A careful inspection of this expression can be performed by
dividing the integration interval into three parts, $\left[  \frac{2\pi}%
{d},\sqrt{\frac{2\pi}{d}\Lambda}\right]  ,$ $\left[  \sqrt{\frac{2\pi}%
{d}\Lambda},\left(  \frac{2\pi}{d}\Lambda^{3}\right)  ^{\frac{1}{4}}\right]  $
and $\left[  \left(  \frac{2\pi}{d}\Lambda^{3}\right)  ^{\frac{1}{4}}%
,\Lambda\right]  $. It turns out that the contributions of the two first intervals
are negligible with respect to the last one. Moreover, checking that in the
range of integration considered, the numerator $q^{2}-\frac{2\pi}{d}\Lambda$
can be approximated by $q^{2}$, we can thus write $\beta \delta F$ for large $d$
as:
\begin{equation}
\beta\delta F\simeq-\frac{1}{d}\frac{L}{2\pi}\int_{0}^{\Lambda}\frac
{\frac{X\left(  0\right)  d}{\pi}\lambda^{2}\Lambda\frac{1}{\left(
\Lambda^{2}+q^{2}\right)  }}{1+\frac{\lambda^{2}}{\pi}\frac{X\left(  0\right)
d}{q}\arctan\left(  \frac{\Lambda}{q}\right)  } \romd q \label{deltafb}
\; .
\end{equation}
Ultimatly, the evaluation of $\beta \delta F$ requires the computation of
\begin{equation}
X ( 0 )  =\TENX( 0,0,0 )  =\frac{1}{d}\int_{-\frac{d}{2}%
}^{\frac{d}{2}}\left[  \frac{1}{2}\dot{\psi}^{4}+\frac{3}{2}\left(  \ddot
{\psi}^{2}+\frac{1}{\lambda^{2}}\dot{\psi}^{2}\right)  \right]  \romd s 
\; .
\label{X0}%
\end{equation}
This computation relies on the saddle point solution $\psi (s)$.
Actually, for large $d$, $\eta\simeq1$ one has%
\[
\dot{\psi} \simeq\sqrt{2\left(  \frac{1-\cos\left(  \psi\right)  }{\lambda
^{2}}\right)  }=\frac{2}{\lambda}\left\vert \sin\left(  \frac{\psi}{2}\right)
\right\vert
\]
whose solution is:
\[
\psi (s)  =2\arccos\left[  \frac{1-C\exp\big(  2\frac{s-s_{0}%
}{\lambda}\big)  }{1+C\exp{\big(2\frac{s-s_{0}}{\lambda}\big)}  }\right]
\]
with $C=\tan^{2}\left(  \frac{\psi_{0}}{4}\right)  $. As a consequence,
replacing $\psi (s)$ in Eq.~(\ref{X0}) one finds directly that $X (0)  =\frac
{16}{3}C\frac{\left(  2C+3\right)  \left(  5C+3\right)  }{d\lambda^{3}\left(
C+1\right)  ^{3}}$. Inserting this value in the expression of $\beta\delta F$ in 
Eq.~(\ref{deltafb}) leads to the result claimed in the text.

\end{appendix}



\begin{thebibliography}{99}
\bibitem {FootnoteG}
The Jacobian is an infinite constant that does not affect
the physical quantities and can thus be disregarded.
\end{thebibliography}
\end{document}